# The Bundles of Intercrossing Fibers of the Extensor Mechanism of the Fingers Greatly Influence the Transmission of Muscle Forces

Anton A Dogadov[1*], Francisco J Valero-Cuevas[2], Christine Serviere[1], Franck Quaine[1*]

[1] Univ. Grenoble Alpes, CNRS, Grenoble INP[**], GIPSA-lab, 38000 Grenoble, France

[2] Alfred E. Mann Department of Biomedical Engineering, and Division of Biokinesiology & Physical Therapy, University of Southern California, Los Angeles, USA

[*]Correspondence: mail@dogadov.fr, franck.quaine@grenoble-inp.fr

[**]Institute of Engineering Univ. Grenoble Alpes

**Abstract:** The extensor mechanism is a tendinous structure that plays an important role in finger function. It transmits forces from several intrinsic and extrinsic muscles to multiple bony attachments along the finger via sheets of collagen fibers. The most important attachments are located at the base of the middle and distal phalanges. How the forces from the muscles contribute to the forces at the attachment points, however, is not fully known. In addition to the well-accepted extensor medial and interosseous lateral bands of the extensor mechanism, there exist two layers of intercrossing fiber bundles (superficial interosseous medial fiber layer and deeper extensor lateral fiber layer), connecting them. In contrast to its common idealization as a minimal network of distinct strings, we built a numerical model consisting of fiber bundles to evaluate the role of multiple intercrossing fiber bundles in the production of static finger forces. We compared this more detailed model of the extensor mechanism to the idealized minimal network that only includes the extensor medial and interosseous lateral bands. We find that including bundles of intercrossing fiber bundles significantly affects force transmission, which itself depends on finger posture. We conclude that the intercrossing fiber bundles — traditionally left out in prior models since Zancolli's simplification — play an important role in force transmission and variation of the latter with posture.

## INTRODUCTION

The extensor mechanism of the fingers of human and non-human primates is a network of tendinous structures that drapes over the dorsum of the finger bones (Van Zwieten et al., 2013). It transmits forces from several extrinsic and intrinsic hand muscles to the phalanges to produce torques at the finger joints (Landsmeer, 1949). This structure plays an important role in finger function, and its disruption degrades manipulation ability. Therefore, it is usually included in detailed biomechanical models of the fingers (Hu et al., 2014; Jadelis et al., 2023; Sachdeva et al., 2015; Francisco J. Valero-Cuevas et al., 2007; Vaz et al., 2015). Even though the extensor mechanism is, in reality, a sheet of intersecting fibers, it has often been idealized as a sparse network of strings (Chao et al., 1989; Garcia-Elias et al., 1991; Schultz et al., 1981; Francisco J.

Valero-Cuevas et al., 2007; Zancolli, 1979). However, the extensor mechanism is a sophisticated continuous fibrous composite structure that can be simplified as having

1. An extensor medial band (central band or slip), which originates from the extrinsic *extensor digitorum communis* muscle and has its principal bone insertion at the proximal part of the middle phalanx as the medial tendon, with middle phalanx attachment;
2. Two interosseous (or intrinsic) lateral bands, radial and ulnar, which originate from the intrinsic muscles. The radial and ulnar bands combine and insert to the proximal part of the distal phalanx as the terminal tendon (Harris and Rutledge, 1972) with distal phalanx attachment;
3. The intercrossing fiber bundles and the extensor hood, connecting the interosseous lateral bands with the extensor medial one (Schultz et al., 1981) The intercrossing fiber bundles are represented by two layers of fibers: interosseous medial fibers and the extensor lateral fibers.

The intercrossing fiber bundles and the extensor hood are of particular interest because they biomechanically couple the forces in the middle and distal phalanx attachments and the rotations of both interphalangeal joints (Leijnse and Spoor, 2012). Moreover, the intercrossing fiber bundles may become more tight or slack as a function of the posture (Leijnse and Spoor, 2012), making the force transmission among the extensor mechanism bands posture dependent (Lee et al., 2008; Sarrafian et al., 1970). This biomechanical coupling has been interpreted as also enabling a nonlinear transmission of tendon forces (i.e., a "switch" behavior) that improves controllability under the anatomical constraints that the fingers do not have any muscles in them (Valero-Cuevas et al., 2007). This means that changing the ratio between the input forces from the intrinsic and extrinsic muscles itself changes the distribution of forces across the middle and distal phalanx attachments. However, we lack detailed studies identifying the posture-dependent interactions by which the multiple fiber bundles of the extensor mechanism enables finger function.

The purpose of this study is to fill this gap in understanding by using a more detailed model of the fiber bundles of the extensor mechanism to understand the role of the extensor hood and the intercrossing fiber bundles on muscle force transmission to produce static fingertip force. In the current study, we focus, without loss of generality, on the extensor mechanism of the middle finger. Applied to the middle finger, the intrinsic muscles, mentioned above, are the second and the third dorsal interosseous muscles, and the second lumbrical muscle. In particular, we built and compared two three-dimensional models of the extensor mechanism: a more detailed model that includes the intercrossing fiber bundles and an extensor hood, and a trivial model, without any structures connecting the extensor medial band with the interosseous lateral bands. We call it the "trivial" model because it reflects the theoretical baseline architecture of muscles where tendons originate in a muscle and insert into bone. While we do not endorse such a trivial structure, this trivial model is not a straw man. Rather, it is the baseline musculotendon anatomy, which evolutionary pressures—presumably of biomechanical nature—drove to specialize into an extensor mechanism. As such, it does help highlight and quantify the biomechanical benefits of a sophisticated extensor mechanism where tendons that originate in muscle combine with other tendons to then insert into bone.

Our results demonstrate changes in force transmission with changes in posture, introduced by the extensor hood and the intercrossing fiber bundles. The functional differences compared to the trivial model speaks to the evolutionary pressures that may have driven the evolution of the topology of the extensor mechanism in the first place, given the anatomical constraints that the fingers do not have any muscles in them and must be actuated by muscles in the palm and forearm. Our model simulating muscle force transmission via bundles of intercrossing fiber bundles now allows us to better understand neuromuscular strategies for finger control, and explain the functional deficits associated with clinically common ruptures or adhesions of the elements of the extensor mechanism. It also enables the design of prostheses and robotics hands using such interconnected tendon architectures.

## METHODS

We coded a custom numerical environment that allows representing the extensor mechanism bands and fiber bundles as a set of strings. Each string consists of a sequence of points, pairwise connected by elastic elements with a linear stress-strain model. This computational environment is written using Matlab 2015 and C++, and is based on the extensor mechanism simulator, described in detail elsewhere (Dogadov et al., 2017). This environment allows simulating tendinous structures with arbitrary topologies and finger postures for static analysis. For a given vector of input muscle forces, it calculates the resulting net joint torques and fingertip wrench (endpoint forces and torques).

The first model (Fig. 1a) was a full extensor mechanism model that includes multiple bundles of intercrossing fiber bundles to approximate the known anatomical bands and sheets of collagenous tissue. The model contains extensor medial band (5), connecting the extrinsic *extensor digitorum* muscle with the medial tendon and middle phalanx attachment (6) of the extensor mechanism to the skeleton. The model also contains the interosseous lateral (or interosseous) bands (4), connecting the intrinsic muscles with terminal tendon and distal phalanx attachment (9). The attachment points of the tendons and ligaments to bones are shown by circles. Finally, the model contains the structures connecting the interosseous lateral band with the medial one. These structures are the extensor hood (1) and the intercrossing fiber bundles: the interosseus medial fibers (2, shown in red) and the extensor lateral fibers (3, shown in blue). These bundles are shown enlarged in Fig. 1d.

The second model was the baseline (trivial) one (Fig. 1b), with no structures, connecting the lateral tendons with the medial one, *i.e.,* it does not contain the extensor hood and intercrossing fiber bundles. The transverse retinacular ligament (7) and triangular ligament (8) were included to both models as they are needed to maintain tendon alignment and prevent bowstringing during force transmission.

Finally, Fig. 1c shows the tendons from the flexor muscles, *flexor digitorum superficialis* (FDS) *and flexor digitorum profundus,* (FDP). Both the full and trivial models included the same tendons from the flexor muscles. We do not include any connection between the flexor tendons and the extensor mechanism.

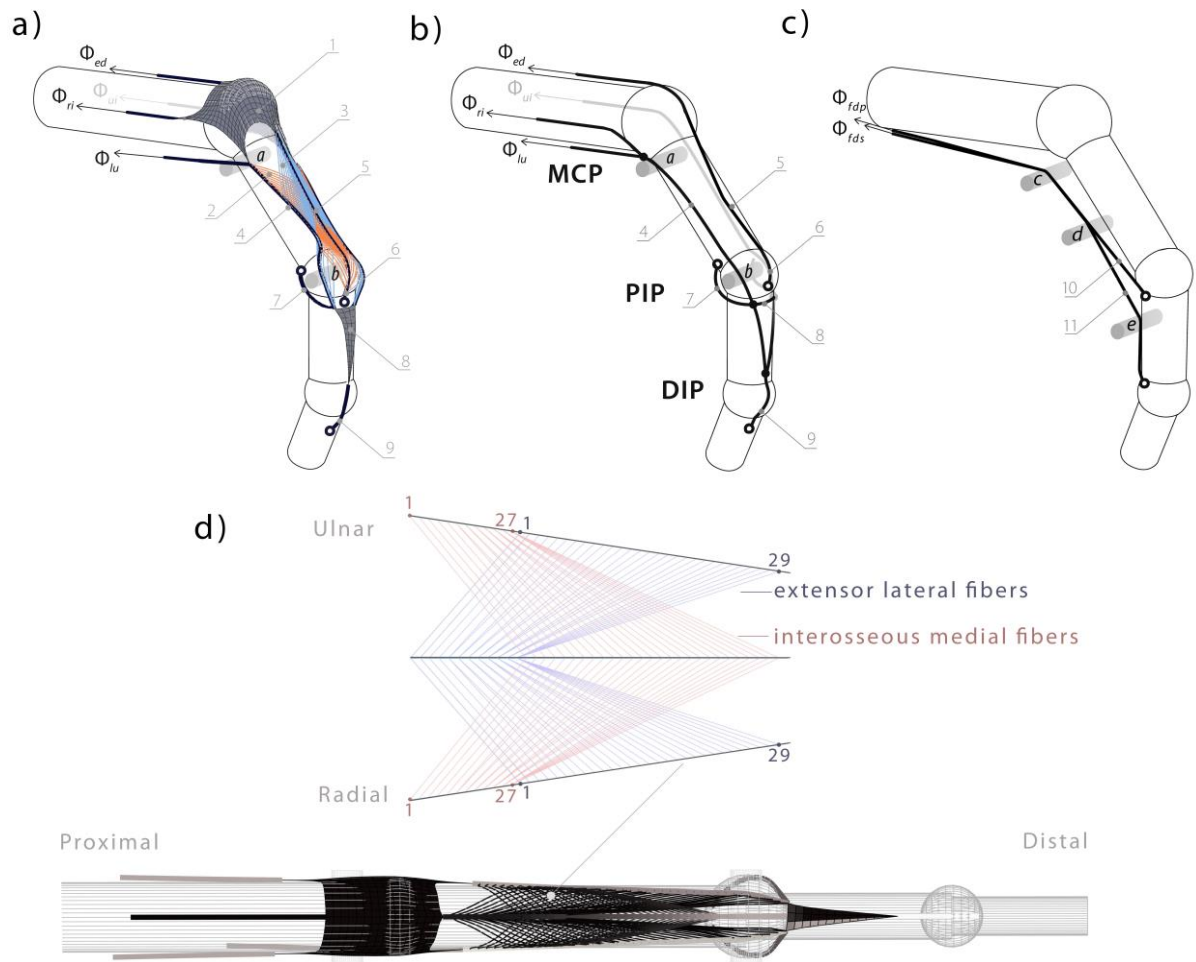

*Fig. 1. The view of the extensor mechanism modelled in a developed environment.*
*a: the full model, which contains the principal tendon and ligaments of the extensor mechanism: 1 – the extensor hood, 2 – interosseous medial fibers (red), 3 – the extensor lateral fibers (blue), 4 – interosseous lateral band, 5 – extensor medial band, 6 – medial tendon and middle phalanx attachment, 7 – transverse retinacular ligament, 8 – triangular ligament, 9 – terminal tendon and distal phalanx attachment;*
*b: the trivial model. The trivial model does not contain the structures connecting the interosseous lateral bands (4) with the extensor medial band (5);*
*c: flexor tendons: 10 – flexor digitorum superficialis tendon, 11 – flexor digitorum profundus tendon (same for both models);*
*d: The schematic view of the intercrossing fiber bundles. Red: interosseous medial fibers; blue: extensor lateral fibers.*

Each extensor mechanism model was draped over on the finger bones in an initial configuration according to anatomical data (Garcia-Elias et al., 1991). The model of the bony anatomy included the metacarpal bone, proximal, middle and distal phalanx of the middle finger. The finger joints considered in the model are a metacarpal (MCP; flexion-extension and ad-abduction), proximal interphalangeal (PIP; flexion-extension), and distal interphalangeal (DIP; flexion-extension) joints. The bones were represented as ideal cylinders capped by spheres. The geometric parameters of the cylinders and spheres were based on anatomical surveys (Buchholz et al., 1992; Darowish et al., 2015) to be, respectively: cylinder lengths 64.6 mm, 44.6 mm, 26.3 mm, 17.4 mm; cylinder radii 4.5 mm, 4.0 mm, 3.0 mm, 2.5 mm; the sphere radii 5.0 mm, 5.4 mm, 4 mm for both models.

In addition to bones, five cylinders (a-e in Fig. 1) with smaller radii were included to the model to avoid tendon bowstringing. The cylinder a is perpendicular to the metacarpal bone and replaces a presumed function of the lumbrical muscle pulley (Stack, 1963); the cylinder b is perpendicular to proximal phalanx bone and replaces the presumed function of the protuberances of $p_1$ head. Cylinders *c, d, e* simulate the annular pulleys that prevent bowstringing of the flexor tendons.

The force of the *extensor digitorum communis* muscle (EDC), ulnar and radial *interosseous* muscle (UI, RI), and lumbrical muscle (LU) were applied to the extensor mechanism model as the input forces. We will note the muscle force values as vector **Φ**:

$$\mathbf{\Phi} = \begin{bmatrix} \Phi_{ed} & \Phi_{ui} & \Phi_{ri} & \Phi_{lu} \end{bmatrix}^T.$$

The deformation of the extensor mechanism due to the applied forces and geometric constraints imposed by the bones and the cylinders *a* and *b* was performed to minimize the overall potential energy (i.e., strain as in (Valero-Cuevas and Lipson, 2004)) of all elastic elements by a Limited memory BFGS algorithm (Liu and Nocedal, 1989; Nocedal, 1980) until the equilibrium state was found, as described in (Dogadov et al., 2017).

Once the equilibrium state of the extensor mechanism was found for a set of applied forces, the tendon tensions internal to the extensor mechanism and resulting force at the attachments were read out. The tensions for each element of the deformed extensor mechanism were found by multiplying its elongation by its stiffness. The forces, transmitted from the extensor mechanism to the bones, including the forces in tendinous attachments and contact forces (the reaction forces created by the tendons overlapping the bones), are used to calculate net joint torques. The torque created by the extensor mechanism were calculated at each kinematic degree of freedom (two for MCP and one each for PIP and PIP). The output fingertip force was found as a product of the finger Jacobian inverse transpose, defined by the finger geometry and a posture, with the joint torque vector. This approach is explained in (Valero-Cuevas, 2015; Valero-Cuevas et al., 1998).

Firstly, we initialized the model with Young's modulus of individual bands and intercrossing fibers in the range of 65–157 MPa and cross-sectional areas of 0.3–1.8 mm² for individual bands and 0.01 mm² for fibers inside bundles. Secondly, we adjusted the stiffness and length of the tendons, as well as the radius of cylinder b (Fig. 1), to match the strain in the extensor medial band, ulnar interosseous lateral band, and terminal tendon to the experimental data reported by (Hurlbut and Adams, 1995) using a particle swarm optimization algorithm (Kennedy and Eberhart, 1995) was used (35 particles, 15 iterations). The root mean squared error between the strain predicted by the model with identified parameters and the experimental data was 0.0125. The detailed comparison between the parametrized model and experimental data is provided in the supporting materials.

Secondly, after parametric identification, this model was used to study the influence of posture on force in intercrossing fibers. To ensure that the observed effect was not dependent on a specific parameter set of the model or the skeleton, we created 60 parameter sets for the extensor mechanism, where the length and elasticity of the bands were modified by applying randomly and uniformly distributed variations (±15 MPa, ±1.5 mm) to identified parameters values. Random variations were also applied to bone elements (length: ±1.5 mm, radii: ±0.5 mm). Each parameter set was used to perform a simulation in three postures (extension MCP = 10°, PIP = 10°, DIP = 10°, mid-flexion MCP = 45°, PIP = 45°, DIP = 10°, and flexion MCP = 90°, PIP = 90°, DIP = 80°) and for two loading conditions: all muscle forces (UI, EDC, RI, and LU) were set to $\Phi$ = 2.9 N and $\Phi$ = 5.9 N (corresponding to 300- and 600-gram loadings, respectively). A two-way ANOVA test was performed to examine the effect of finger posture on force distribution in the intercrossing fiber bundles. For each of the 60 parameter sets we calculated the mean force across all fibers within the same bundle (ulnar interosseous medial fibers, ulnar extensor lateral fibers, radial interosseous medial fibers, radial extensor lateral fibers). The mean force in a bundle served as the dependent variable, while bundle type and posture were the two independent variables. The test was repeated for both loading conditions.

Finally, we compared the finger model with the full extensor mechanism with identified parameters with the finger model with trivial extensor mechanism across three postures and two loading conditions. We studied the feasible tendon force set, which is a set of all possible combinations of the forces that can be transmitted by the full or trivial extensor mechanism model to the medial tendon (middle phalanx attachment) and the terminal tendon (distal phalanx attachment) for all possible combination of muscle forces (UI, EDC, RI, and LU getting values either 0 or $\Phi$, where $\Phi$ was 2.9 N or 5.9 N according to loading condition). A feasible tendon force set was calculated as a convex hull in a plane of magnitude of medial and terminal tendon forces for 16 possible combinations of four muscle forces $\left( \Phi = [0\ 0\ 0\ 0]^T, [\Phi\ 0\ 0\ 0]^T, ..., [\Phi\ \Phi\ \Phi\ \Phi]^T \right)$. Similarly, feasible force set at the fingertip was calculated for three postures and two loading conditions to compare a full and a trivial extensor mechanism models.

# RESULTS

Fig. 2 shows the force distribution among the intercrossing fiber bundles (ulnar side) across postures, based on 60 extensor mechanism simulations with randomly varied parameters. The fiber numbers in Fig. 2 are the same as in Fig. 1d. The simulation results for fiber bundles from both sides of the finger are shown in supporting materials.

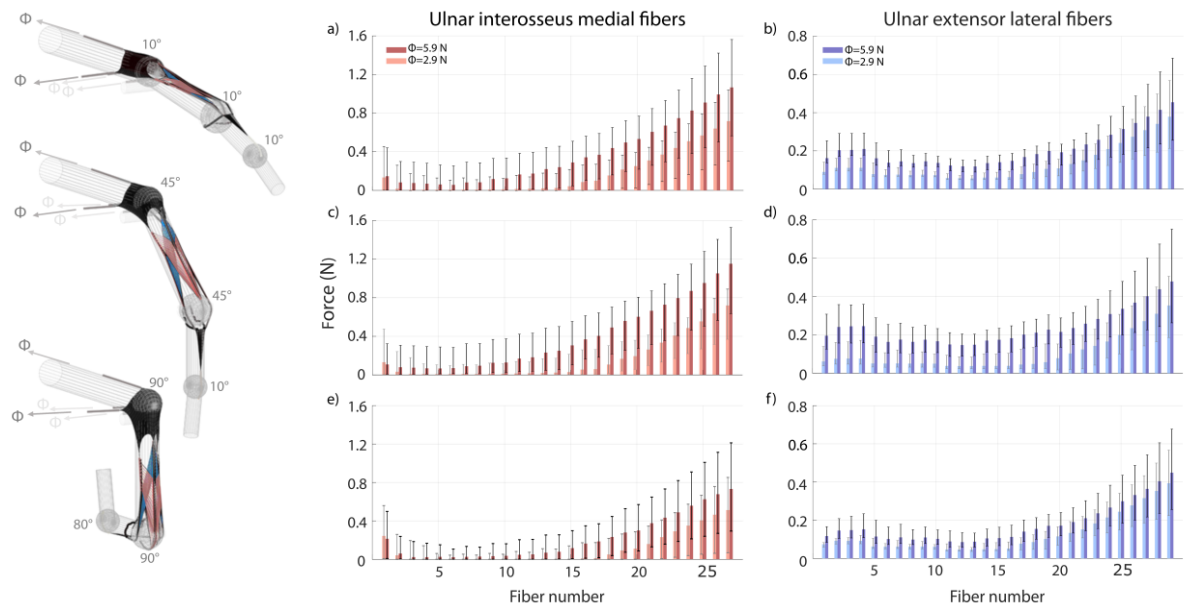

*Fig. 2. The influence of posture on the forces in ulnar intercrossing fiber bundles. Red: ulnar interosseous medial fibers; blue: extensor lateral fibers. The first row (pannels a and b) corresponds to extension (MCP = 10°, PIP = 10°, DIP = 10°), the second row corresponds to mid-flexion (MCP = 45°, PIP = 45°, DIP = 10°, pannels c and d), and the third row corresponds to flexion (MCP = 90°, PIP = 90°, DIP = 80°, pannels e and f). Two loading conditions are shown by bars of light (Φ = 2.9 N) and dark (Φ = 5.9 N in UI, EDC, RI, and LU muscles) colors, respectively. Within each loading condition, the same force magnitude was applied to all muscles. The median of 60 simulations, along with the 5th and 95th percentiles, is shown.*

The two-way ANOVA showed that the effect of posture on forces in intercrossing fiber bundles for both loading conditions (2.9 N loading: $F(2,708)=32.56$, $p<0.001$, 5.9 N loading: $F(2,708)=129.84$, $p<0.001$). There was a significant posture × fiber bundle interaction effect (2.9 N loading $F(6,708)=13.54$, $p<0.001$), 5.9 N loading $F(6,708)=15.52$) suggesting that the influence of posture varies according the bundles.

Fig. 3 shows the changes in the feasible tendon force set of the full extensor mechanism model with identified parameters (left column) in comparison with a trivial model (right column). The full-loading state, which was the state when all four extensor muscle forces were equal to Φ, is shown by a circle in each panel.

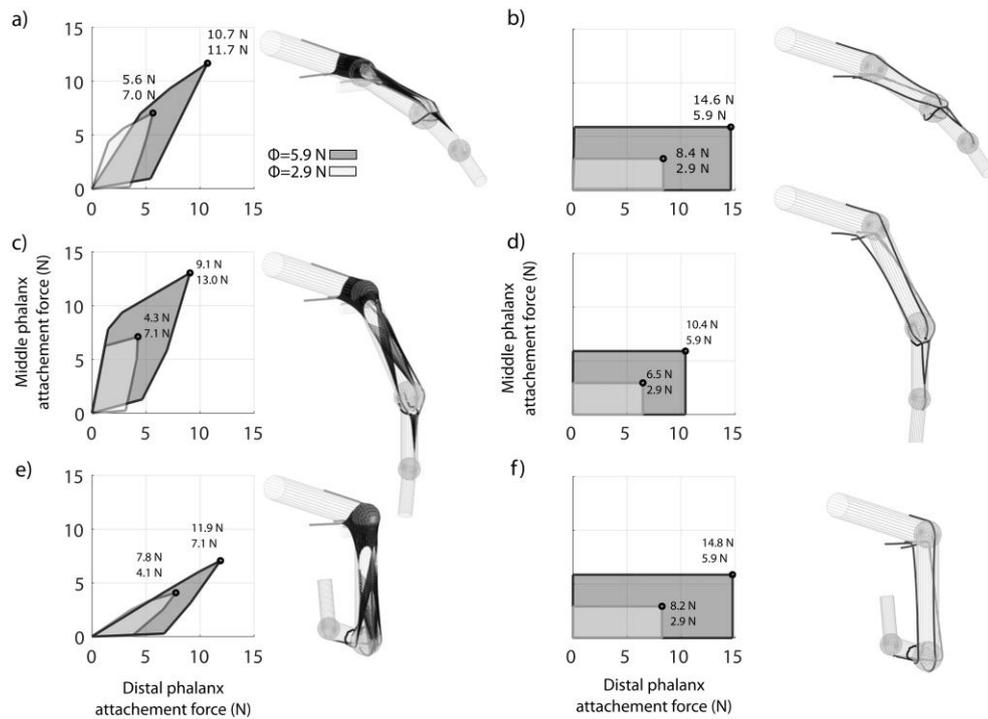

Fig. 3. The effect of the posture on feasible tendon force set. Left column corresponds to a full extensor mechanism model, right collumn corresponds to a trivial model. The first row corresponds to extension (MCP = 10°; PIP = 10°; DIP = 10°), the second row coresponds to mid-flexion (MCP = 45°; PIP = 45°; DIP = 10°), and the third row corresponds to flexion posture (MCP = 90°; PIP = 90°; DIP = 80°). The full-loading state, which corresponds to loading of the extensor mechanism models by all four muscles, is shown by a circle in each feasible tendon force set. The middle and distal phalanx attachment force values in full-loading sate are comparable for both models, but the areas of the feasible tendon force set are smaller for the full model. Also for a full model, the shape and orientation of the feasible tendon force set change with posture. Dark area cooresponds to 5.9 N loading and light area correponds to 2.9 N loading.

It can be seen from the right column of the image that the feasible tendon force set of the trivial model had a rectangular shape for all postures. The maximal force in the middle phalanx attachment did not change with posture and remained at 2.9/5.9 N, depending on the loading condition (2.9/5.9 N loading). The maximal force in the distal phalanx attachment was similar in extension (8.4/14.8 N) and flexion (8.2/14.6 N) but decreased in mid-flexion (6.5/10.4 N). This may be explained by the fact that the force in the distal attachment is controlled by the interosseous lateral bands, which are connected by a triangular membrane, while stretching of the triangular membrane in flexion may influence the force at the distal phalanx attachment.

The ratio between the force in the middle and distal phalanx attachment of the trivial extensor mechanism model in the full-loading state was 0.35/0.40 in extension, 0.45/0.57 in mid-flexion, and 0.35/0.40 in flexion.

Similar to the trivial model, the force in the distal phalanx attachment reached a minimum value in mid-flexion. However, contrary to the trivial model, the shape, size, and orientation of the feasible tendon force set in the full model changed with posture. As a result, the force in the middle attachment also varied differently with posture.

It can also be observed that the area of the feasible tendon force set for the full model was smaller than the corresponding areas for the trivial model in extension and flexion postures (-18% for extension posture, -55% for flexion posture at 2.9 N loading; -55% and -72% at 5.9 N loading).

Fig. 4 shows the effects of the posture on sagital plane projections of the feasible fingertip force set (FFS). The left column corresponds to the full extensor mechanism model with identified parameters, the right column to the trivial extensor mechanism model. The full-loading fingertip force, which was produced by the model when all four extensor muscles forces were equal to $\Phi=2.9/5.9$ N, is shown by a circle in each panel ("full loading").

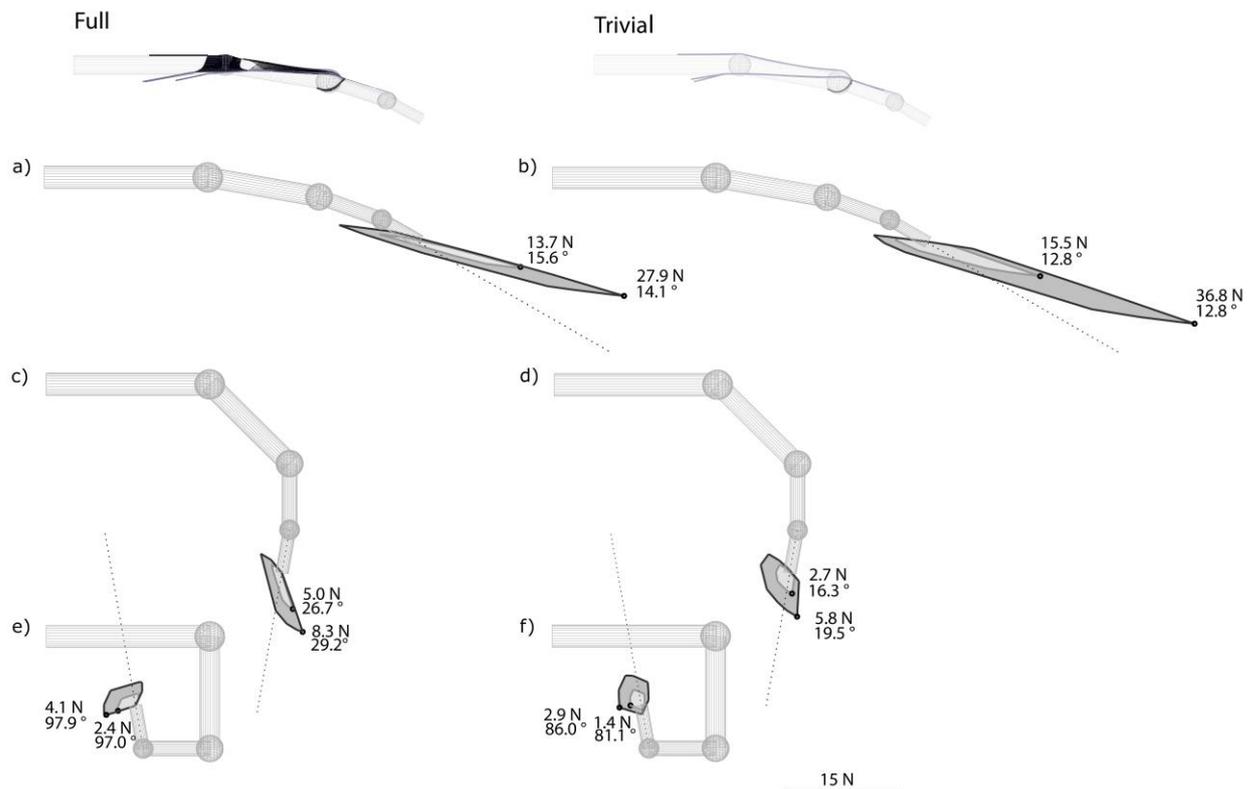

*Fig. 4. Influence of the posture on sagital plane projection of the feasible force set. Left column corresponds to a full extensor mechanism model, right collumn corresponds to a trivial model. First row corresponds to extension posture (MCP = 10°; PIP ;= 10°; DIP = 10°), second row coresponds to mid-flexion posture (MCP = 45°; PIP = 45°; DIP = 10°), third row corresponds to flexion posture (MCP = 90°; PIP = 90°; DIP = 80°). Dark area cooresponds to 5.9 N loading and light area correponds to 2.9 N loading.*

For both trivial and full model, the changes of shape and orientation of FFS with posture were observed. The angle between the distal phalanx and the sagital plane projection of the full-loading force was higher in full model than in trivial one (for 2.9 N loading condition, mid-flexion posture: 26.7° for full model vs 16.3° for trivial model). Finally, it can also be observed from the figure that the area of the FFS in the full extensor mechanism model was smaller than that of the trivial model in the extension posture (FFS area of the full model is 55% that of the trivial model for both 2.9 N and 5.9 N loadings).

## DISCUSSION

We demonstrated that the intercrossing fiber bundles and the extensor hood reduces the area of feasible tendon force set. The full extensor mechanism model, which contain the intercrossing fiber bundles and the extensor hood, is lower than the areas of feasible tendon force set and FFSs, produced by the trivial model, in which there are no connections between the extensor medial and interosseous lateral bands. This area increases due to the fact that the trivial extensor mechanism model enables the independent control of the forces in the middle and distal phalanx attachment. However, in the case of the full model of the extensor mechanism, these forces are naturally coupled.

Secondly, we have shown that the bundles of intercrossing fiber can modify the force distribution according to posture. This may indicate that the nervous system has to modulate the sharing in involved muscle and intensity according to the finger posture in order to produce the wanted fingertip force. This implies that there exists a link between the passive adaptations of the extensor mechanisms and the active modulation of the muscle recruitments for useful fingertip tasks, such as grasping objects (Wei et al., 2022), writing (Gerth and Festman, 2023), or playing musical instrument (Furuya et al., 2011).

The analyzed full model has several limitations. Firstly, the model topology oversimplifies the real extensor mechanism anatomy. Over MCP joint the extensor mechanism was represented only by the extensor hood. However, the metacarpophalangeal fibrous griddle, or sagittal band, which connect the extensor tendons to the deep transverse intermetacarpal ligament and capsular join (Zancolli, 1979) was not taken into account. Moreover, no attachments of the extensor mechanism at the base of the proximal phalanx were taken into account. Secondly, the bones were modeled as cylinders with spheres corresponding to the joints.

This study is limited in that it does not include all other muscles acting on the finger, but this work enables future work to understand the function of the human fingers that considers their complex anatomy in more detail.

In addition, this work only considered force transmission by the trivial model, but does not consider other important biomechanical consequences of it. First and foremost is the need to maintain and regulate the tendon path as the finger changes posture, where the "unsupported" trivial tendons may slide, bowstring, cause rapid changes in moment arms and even cause tendinitis or scaring during their unguided sliding movement. In our model, the path of the tendons in the trivial model was enforced arbitrarily. From this perspective, the extensor mechanism may server to retain force transmission while also serving as a support and guiding structure, much like the annular bands and sesamoids in other tendons.

And secondly, there are other considerations in addition to tendon force and joint torque production. Recent work has suggested that tendon force transmission is important for other important aspects of function such as stability during force production (Sharma and Venkadesan, 2022). Similarly, producing slow finger movements very likely depend more on managing the internal strain energy of the system and not second-order rigid-body dynamics driven by joint torques or muscle forces (Babikian et al., 2016).

As such, the evolutionary pressures for the formation of the extensor mechanism may not be strictly limited to force transmission. That is, the extensor mechanisms may have been a multi-factorial evolutionary adaptation that also allows for stability and accurate slow movements with the fingertips that gave human primates and also non-human catarrhine primates a competitive advantage for effective manipulation capabilities.

## ACKNOWLEDGEMENTS

The work was supported by IDEX Université Grenoble Alpes scholarship for international mobility. The authors acknowledge Vishweshwer Shastri (USC) and Gelu Ionescu (GIPSA-Lab) for their assistance with the programming of the extensor mechanism simulator.

## ETHICAL STATEMENT

None.